\documentclass[12pt,eqsecnum]{revtex4}
\begin{document}

 \title{Field dependent nilpotent symmetry for gauge theories}

\author{ Sudhaker Upadhyay\footnote {e-mail address: sudhakerupadhyay@gmail.com}}
\author{ Bhabani Prasad Mandal\footnote{e-mail address:
bhabani.mandal@gmail.com}}

\affiliation { Department of Physics, \\
Banaras Hindu University, \\
Varanasi-221005, INDIA.  }
 
\begin{abstract}
We construct the field dependent  mixed BRST (combination of BRST and anti-BRST) transformations for
pure gauge theories. These are shown to be an exact nilpotent symmetry of  both the effective action as well as
the generating functional for certain choices of the field dependent parameters. We show that the Jacobian 
contributions
for path integral measure in the definition of generating functional arising from BRST and anti-BRST part 
compensate each other.
The field dependent mixed BRST transformations are also considered in field/antifield formulation to show that 
the solutions of quantum master equation remain invariant under these. 
Our results are supported by several explicit examples. 
\end{abstract}

 \maketitle

\section{Introduction}
The BRST transformation plays a central role in  
the quantization, renormalizability, unitarity and 
other aspects of the gauge
theories \cite{brst,tyu,ht,wei,chi}. 
Such nilpotent transformation, characterized by an infinitesimal,
global   and
anticommuting parameter, leaves the effective action including the gauge fixing term  invariant.
Similar to the BRST transformation, anti-BRST transformation is also a symmetry transformation  where the role of
ghost and anti-ghost fields are interchanged. The anti-BRST symmetry does not play as fundamental role as 
BRST symmetry itself but it is a useful tool in geometrical description   of BRST/anti-BRST symmetric theories 
\cite{boto} and widely  used in the investigation of perturbative renormalization of the gauge  
theories 
\cite{gaba}.  We consider the mixed BRST (MBRST) transformation (defined as 
$\delta_m \phi =s_b\phi\ \delta\Lambda_1 +s_{ab}\phi\ \delta\Lambda_2$) with infinitesimal, anticommuting
but global parameters   $\delta\Lambda_1$ and $\delta\Lambda_2$  corresponding
to the BRST variation ($s_b$) and anti-BRST variation ($s_{ab}$) of the generic fields  $\phi$ respectively. However,
the transformations $s_b$ and $s_{ab}$  defined in the MBRST transformation satisfy the absolute
anticommuting relation $\{s_b, 
s_{ab}\}\phi=0$.
Such an infinitesimal 
mixed transformation is also a
nilpotent symmetry transformation of the effective action as well as of the generating functional. 

In the present  work we construct the field dependent MBRST (FMBRST) transformation having finite and field dependent 
parameters. The usual finite field dependent BRST (FFBRST) and anti-BRST (FF-anti-BRST) transformations 
are the symmetry transformations of  the effective action only but do not leave the generating functional 
invariant as the path integral measure in the definition of generating functional transforms in a nontrivial 
manner \cite{jm,susk}. The FFBRST and FF-anti-BRST transformations for the effective theory have many 
applications in gauge field theory \cite{jm,susk,srbpm,cou,rb,subp,ssb,subp1,subp2}.
 Unlike the usual FFBRST and FF-anti-BRST
transformations the  FMBRST transformations  are shown to be  the symmetry of the 
both  effective 
action as well as the generating functional of the theory. We construct the  finite parameters in the FMBRST 
transformation in 
such a way that the Jacobian contribution due to FFBRST part compensates the same due to FF-anti-BRST 
part. 
Thus we are able to construct the finite nilpotent transformation which leaves the generating 
functional as well as the effective action of the theory invariant. We further show that the effect of FMBRST 
transformation is equivalent to the effect of successive operations of FFBRST and FF-anti-BRST transformations. 

Our results are supported by several explicit examples. First of all 
we consider the gauge  invariant model for single 
self-dual chiral boson in (1+1) dimensions \cite{pp,pp1,sud}, which is very useful in the study of certain string 
theoretic models \cite{ms,ai} and plays a very crucial role in the study of quantum Hall effect \cite{wen}.
(3+1) dimensional Abelian as well as  non-Abelian Yang-Mills (YM) theory in the Curci-Ferrari-Delbourgo-Jarvis (CFDJ) 
gauge 
\cite{cf,cf1,del} are also considered to demonstrate the above
finite nilpotent symmetry.

The Lagrangian quantization of Batalin and Vilkovisky (BV) formulation \cite{bv1,bv2,ht,wei}, which is also known as 
the field/antifield   formulation, is considered to be one of the most powerful and advanced technique
of quantization of gauge theories involving the BRST symmetry \cite{ht,wei,sudbpm}. To study the 
role of FMBRST transformation in field/antifield formulation we consider the same 
 three  simple models in BV formulation. We show that the FMBRST transformation does not 
change  the generating functional written in terms of extended quantum action in BV formulation. Hence
the FMBRST 
transformation leaves the different solutions of 
the quantum master equation in field/antifield formulation invariant. 

The paper is organized as follows. In Sec. II, we discuss the infinitesimal MBRST
transformation. Next the FMBRST transformation is constructed in Sec. III.
The non-trivial Jacobian  for such FMBRST 
transformation is evaluated in Sec. IV. Sec. V, is devoted to the explicit examples having FMBRST symmetry 
transformation.
In Sec. VI, we study the field/antifield formulation in the context of FMBRST transformation.
Sec. VII is reserved for concluding remarks.

\section{The infinitesimal  MBRST transformation}
The generating functional for the Green's function in an effective theory described by the effective action $S_{eff} 
[\phi]$  is defined as
\begin{equation}
Z=\int  D\phi\ e^{iS_{eff}[\phi]},
\end{equation}
 \begin{equation}
S_{eff} [\phi]= S_0[\phi] +S_{gf}[\phi] +S_{gh}[\phi],
\end{equation} 
where  $\phi$ is the generic notation for all fields 
 involved in the effective theory.
The infinitesimal  
 BRST ($\delta_b$) and anti-BRST 
($\delta_{ab}$) transformations are defined as
\begin{equation}
\delta_b \phi =s_b\phi\ \delta\Lambda_1,\ \ s_b^2=0
\end{equation}
\begin{equation}
\delta_{ab} \phi =s_{ab}\phi\ \delta\Lambda_2,\ \ s_{ab}^2=0,
\end{equation}
where $\delta\Lambda_1$ and $\delta\Lambda_2$ are infinitesimal, anticommuting but global parameters.  
 Such transformations leave the generating functional as well as effective action 
invariant 
\begin{equation}
\delta_b Z=0=\delta_b S_{eff},
\end{equation}
\begin{equation}
\delta_{ab} Z =0=\delta_{ab} S_{eff}.
\end{equation}
This implies that the effective action $S_{eff} [\phi]$ and the generating functional are also  invariant 
under the MBRST  ($\delta_m =\delta_b +\delta_{ab}$) transformation 
\begin{equation}
\delta_m Z =0= \delta_m S_{eff}. \label{mix}
\end{equation}
Further such MBRST transformation is nilpotent because,
\begin{equation}
\{ s_b, s_{ab}\} =0.
\end{equation}
Now, in the next section we  construct the finite field dependent version of following infinitesimal MBRST symmetry 
transformation 
  \begin{equation}
\delta_m \phi = s_b\phi\ \delta\Lambda_1 +s_{ab}\phi\ \delta\Lambda_2.\label{infi}
\end{equation}

\section{Construction of FMBRST transformation}
To construct the FMBRST transformation, we use to follow the similar method of 
constructing FFBRST transformation \cite{jm}. However, in this case unlike FFBRST transformation
 we have to deal with two 
parameters, one for 
the BRST transformation and the other for anti-BRST transformation. We introduce a numerical parameter $\kappa 
(0\leq \kappa\leq 1)$ and make all the fields ($\phi (x,\kappa)$) $\kappa$-dependent in such a way that
 $\phi(x,\kappa =0)\equiv \phi(x)$ and $\phi(x,\kappa 
=1)\equiv \phi^\prime(x)$, the transformed field. Further, we make the infinitesimal parameters $\delta\Lambda_1$ and
$\delta\Lambda_2$  field dependent as
\begin{eqnarray}
\delta\Lambda_1&=&\Theta_1'[\phi(x,\kappa)]d\kappa\\
\delta\Lambda_2&=&\Theta_2'[\phi(x,\kappa)]d\kappa,
\end{eqnarray}
 where the prime denotes the derivative with respect to $\kappa$ and 
 $\Theta'_i [\phi(x,\kappa)] (i=1,2)$ are  infinitesimal field dependent parameters.
The infinitesimal but field dependent MBRST transformations, thus can be written generically as 
\begin{eqnarray}
\frac{d\phi(x,\kappa)}{d\kappa}=s_b \phi (x,\kappa ) \Theta_1^\prime [\phi (x,\kappa )] 
+
s_{ab} \phi (x,\kappa ) \Theta_2^\prime [\phi (x,\kappa )]. 
\label{diff}
\end{eqnarray} 
 Following the work in Ref.\cite{jm} it can be shown that the parameters 
 $\Theta_i^\prime[\phi(x, \kappa)] (i =1,2),$ contain  the factors
$\Theta_i^\prime[\phi(x, 0)] $ $(i=1,2),$ which are considered to be nilpotent.
 Thus $\kappa$ dependency from 
$\delta_b\phi(x,\kappa)$ and $\delta_{ab}\phi(x,\kappa)$ can be dropped. 
Then the Eq. (\ref{diff}) can be written as 
\begin{equation}
\frac{d\phi(x,\kappa)}{d\kappa}= s_b \phi (x,0 ) \Theta_1^\prime [\phi (x,\kappa )] +
s_{ab} \phi (x, 0 ) \Theta_2^\prime [\phi (x,\kappa )].
\end{equation}
The FMBRST transformations with the finite field dependent parameters then can be 
constructed by integrating such infinitesimal transformations from $\kappa =0$ 
to $\kappa= 1$, such that
\begin{eqnarray}
\phi^\prime &\equiv & \phi (x,\kappa =1)=\phi(x,\kappa=0)+s_b \phi (x) \Theta_1 [\phi (x
 )]\nonumber\\
 &+& s_{ab} \phi (x) \Theta_2 [\phi (x)],
\label{kdep} 
\end{eqnarray} 
where 
\begin{equation}
\Theta_i[\phi(x)]=\int_0^1 d\kappa^\prime\Theta_i^\prime [\phi(x,\kappa^\prime)],
\end{equation}
 are the finite field dependent parameters with $i=1, 2$.  

Therefore, the FMBRST transformation corresponding MBRST transformation mentioned in Eq. (\ref{infi})
is given by
\begin{equation}
\delta_m \phi = s_b\phi\ \Theta_1 +s_{ab}\phi\ \Theta_2 
\end{equation}
It can be shown that above FMBRST transformation with some specific choices of the finite 
parameters $\Theta_1$ and $\Theta_2$ is the symmetry transformation of the both effective action
and the generating functional as the 
path integral measure is   generally  invariant under such transformation.
\section{Method for Evaluating the  Jacobian} 
For the symmetry of the generating functional we need to calculate the Jacobian of the 
path integral measure in the definition of generating functional.
The Jacobian  of the path integral measure for FMBRST transformation $J  $ can be evaluated for some 
particular choices of the finite field dependent parameters $\Theta_1[\phi(x)]$ and 
$\Theta_2[\phi(x)]$. We start with the definition,
\begin{eqnarray}
D\phi  &=&J( \kappa)\ D\phi(\kappa) \nonumber\\ 
&=& J(\kappa+d\kappa)\ D\phi(\kappa+d\kappa),
\end{eqnarray} 
 Now the transformation from $\phi(\kappa)$ to $\phi(\kappa +d\kappa)$ is infinitesimal 
in nature, thus the infinitesimal change in Jacobian can be calculated as 
\begin{equation}
\frac{J (\kappa)}{J (\kappa +d\kappa)}=\Sigma_\phi\pm 
\frac{\delta\phi(x, \kappa)}{
\delta\phi(x, \kappa+d\kappa)} 
\end{equation}
where $\Sigma_\phi $ sums over all fields involved in the path integral measure  
and  $\pm$ 
sign refers to whether $\phi$ is a bosonic or a fermionic field.
Using the Taylor expansion we calculate the above expression as 
\begin{eqnarray} 
 \frac{1}{J (\kappa)}\frac{dJ (\kappa)}{d\kappa} 
&=& - \int d^4x\left [\Sigma_\phi (\pm) s_b  
\phi (x,\kappa )\frac{
\partial\Theta_1^\prime [\phi (x,\kappa )]}{\partial\phi (x,\kappa )} \right.\nonumber\\
&+&\Sigma_\phi(\pm) \left. s_{ab}  \phi (x,\kappa )\frac{
\partial\Theta_2^\prime [\phi (x,\kappa )]}{\partial\phi (x,\kappa )}\right ].\label{jac}
\end{eqnarray}

The Jacobian, $J(\kappa )$, can be replaced (within the functional integral) as
\begin{equation}
J(\kappa )\rightarrow e^{i (S_1[\phi ] + S_2[\phi ] )}
\end{equation}
iff the following condition is satisfied 
\begin{eqnarray}
\int &D\phi (x)& \left [ \frac{1}{J }\frac{dJ }{d\kappa} -i\frac
{d  S_1[\phi (x,\kappa )]}{d\kappa} -i\frac{d S_2[\phi (x,\kappa )] }{d\kappa}\right ]\nonumber\\
&\times & e^{ i(S_{eff}+S_1+S_2) }=0, 
\label{mcond}
\end{eqnarray} 
where $ S_1[\phi ]$ and  $ S_2[\phi ]$ are some local functionals of fields and satisfy the initial
condition
 \begin{equation}
S_i[\phi(\kappa =0) ]=0,\ i =1, 2.\label{inco}
\end{equation}
The finite parameters $\Theta_1$ and $\Theta_2$ are arbitrary and we can construct them in such a way that
the infinitesimal change in Jacobian $J$ (Eq. (\ref{jac})) with respect to $\kappa$ vanishes  
\begin{equation}
\frac{1}{J}\frac{dJ}{d\kappa}=0.\label{j}
\end{equation}
Therefore, with the help of Eqs. (\ref{mcond}) and (\ref{j}), we see that
\begin{equation}
\frac
{d S_1[\phi (x,\kappa )]}{d\kappa} +\frac{S_2[\phi (x,\kappa )] }{d\kappa}=0.
\end{equation}
It means the $S_1 +S_2$ is independent of $\kappa $ (fields) and must vanish to satisfy the initial condition
 given in Eq. (\ref{inco}) satisfied.
Hence the generating functional is not effected by the Jacobian $J$ as $J =e^{i(S_1+S_2)} = 1$. 
The nontrivial Jacobian arising from finite BRST parameter  $\Theta_1$ compensates the same arising due to finite 
anti-BRST parameter  $\Theta_2$. It is straight forward to see that the  effective action $S_{eff}$ is invariant 
under such FMBRST transformation.
\section{Examples} 
To demonstrate the results obtained in the previous section we would like to consider several explicit examples in 
(1+1) as well as in (3+1) dimensions. In particular we consider the 
bosonized self-dual chiral model in (1+1) dimensions, 
Maxwell's theory in (3+1) dimensions, and non-Abelian YM theory in (3+1) dimensions. In all these cases we construct 
explicit finite parameters $\Theta_1$ and $\Theta_2$ of FMBRST transformation such that
the generating functional  remains invariant.
\subsection{Bosonized chiral model}
We start with the generating functional for the bosonized self-dual chiral model as \cite{pp,pp1}
\begin{equation} 
Z_{CB}=\int D\phi\ e^{iS_{CB}},\label{z}
\end{equation}
where   $D\phi$ is the path integral measure in generic notation. The  effective action $S_{CB}$  in (1+1) 
dimensions is given as
 \begin{eqnarray} 
{S}_{CB}&=& \int d^2x [\pi_\varphi\dot\varphi +\pi_\vartheta\dot\vartheta 
+p_u\dot u-\frac{1}{2}
\pi_\varphi^2
+\frac{1}{2}\pi_\vartheta^2 \nonumber\\
&+&\pi_\vartheta (\varphi'-\vartheta' +\lambda)
+ \pi_\varphi\lambda +\frac{1}{2}B^2 \nonumber\\
&+&B(\dot \lambda -\varphi -\vartheta ) 
 + \dot{\bar c}\dot c
-2\bar c c],
\end{eqnarray}
where the fields $\varphi, \vartheta, u, B, c$ and $\bar c$  are  the self-dual field, Wess-Zumino field,
multiplier field, auxiliary 
field, ghost field and anti-ghost field respectively.
The   nilpotent 
BRST and anti-BRST transformations for this theory are

BRST:
\begin{eqnarray}
\delta_{b}\varphi &=&  c\ \delta \Lambda_1,\ \ \ \delta_{b} \lambda =-\dot { c
}\ \delta \Lambda_1, \ \ \ \delta_{b} \vartheta = c \ \delta \Lambda_1,\nonumber\\
\delta_{b}  \pi_\varphi &=& 0,\ \ \ \delta_{b}  u =0, \ \ \ \delta_{b}  \pi_\vartheta =0,\ \ \ 
\delta_{b} \bar c =  B \ \delta \Lambda_1,\nonumber\\
 \delta_{b}  B &=&0, \ \ \ \delta_{b}   c =0,\ \ \delta_{b}  p_u 
=0,\label{sym}
\end{eqnarray}
anti-BRST: 
\begin{eqnarray}
\delta_{ab}\varphi &=& -\bar c\ \delta \Lambda_2,\ \ \ \delta_{ab}\lambda =\dot {\bar c
}\ \delta \Lambda_2, \ \ \ \delta_{ab}\vartheta = -\bar c\ \delta \Lambda_2,\nonumber\\
\delta_{ab} \pi_\varphi &=& 0,\ \ \ \delta_{ab} u =0, \ \ \ \delta_{ab} \pi_\vartheta =0,\ 
\delta_{ab}c  =  B\ \delta \Lambda_2,\nonumber\\
 \delta_{ab} B &=&0, \ \ \ \delta_{ab} \bar c =0,\ \ \delta_{ab} p_u 
=0,\label{ab}
\end{eqnarray}
where $\delta\Lambda_1$ and $\delta\Lambda_2$  are infinitesimal, anticommuting and global parameters.
Note that $s_{b}$ and $s_{ab}$ are absolutely anticommuting i.e.
$(s_{b}s_{ab}+s_{ab}s_{b})\phi=0$. In this case the  MBRST symmetry transformation
  ($\delta_m\equiv \delta_{b}+\delta_{ab} $), as constructed in section II, reads as  
\begin{eqnarray}
\delta_m\varphi &=&  c\ \delta \Lambda_1 -\bar c\ \delta \Lambda_2,\ \ \ \delta_m\lambda =-\dot { c
}\ \delta \Lambda_1 +\dot {\bar c}\ \delta \Lambda_2,\nonumber\\
 \delta_m\vartheta &=& c \ \delta \Lambda_1 -\bar c\ \delta \Lambda_2,\ \delta_m \pi_\varphi = 0,\ \ \ \delta_m u =0, 
\nonumber\\
 \delta_m \pi_\vartheta &=&0,\ \delta_m\bar c = B \ \delta \Lambda_1,\ \ \ \delta_m B =0,\
\delta_m  c =  B\ \delta \Lambda_2,\nonumber\\ 
 \delta_m p_u &=&0.\label{b}
\end{eqnarray} 

The FMBRST transformations corresponding to the
 above MBRST transformations are constructed as
\begin{eqnarray} 
\delta_m\varphi &=&  c\ \Theta_1 -\bar c \ \Theta_2,\   \delta_m\lambda =-\dot { c
}\ \Theta_1 +\dot {\bar c}\ \Theta_2,\nonumber\\
 \delta_m\vartheta &=& c\ \Theta_1 -\bar c\ \Theta_2,\ 
\delta_m \pi_\varphi = 0,\ \ \ \delta_m u =0,\nonumber\\
 \delta_m \pi_\vartheta &=&0,\ 
\delta_m\bar c  =  B \ \Theta_1,\ \ \ \delta_m B =0, \nonumber\\
 \delta_m  c &=& B\ \Theta_2,\ \ \delta_m p_u 
=0,\label{c}
\end{eqnarray}
where $\Theta_1$ and $\Theta_2$ are finite field dependent parameters and are still 
anticommuting 
in nature. 
 We construct  the finite  parameters $\Theta_1$ and $\Theta_2$ as 
\begin{equation}
\Theta_1= \int\Theta'_1 d\kappa= \gamma\int d\kappa\int d^2x [\bar c(\dot\lambda-\varphi -\vartheta )],\label{fi}
\end{equation}
\begin{equation}
\Theta_2=\int\Theta'_2d\kappa= -\gamma\int d\kappa \int d^2x [ c(\dot\lambda-\varphi -\vartheta )],\label{fia}
\end{equation}
where $\gamma$ is an arbitrary parameter.

Using Eq. (\ref{jac}), the infinitesimal change in Jacobian for the FMBRST transformation given in Eq. (\ref{c})  
can be calculated 
 as 
\begin{equation}
\frac{1}{J}\frac{dJ}{d\kappa} = 0.
\end{equation} 

The contribution from second and third terms in the R.H.S. of Eq. (\ref{jac}) cancels each other.
This implies that the Jacobian for path integral measure is unit under FMBRST transformation. Hence the generating 
functional as well as the effective action are invariant under FMBRST 
transformation
\begin{eqnarray}
Z_{CB} \left(  \int D\phi\ e^{iS_{CB}} \right) \stackrel{FMBRST}{----\longrightarrow }
Z_{CB}.
\end{eqnarray} 

Now, we would like to consider of the effect of FFBRST transformation with finite parameter $\Theta_1$ and 
FF-anti-BRST transformation with finite parameter $\Theta_2$ independently.
The infinitesimal change in Jacobian $J_1$ for the FFBRST  transformation with the parameter $\Theta_1$  
is calculated as
\begin{equation}
\frac{1}{J_1}\frac{dJ_1}{d\kappa} =  \gamma\int d^4x \left[B (\dot\lambda -\varphi -\vartheta) +\dot{\bar c}\dot c -
2\bar c c\right].\label{jac1}
\end{equation}
To write the Jacobian $J_1$ as $e^{iS_1}$ in case of BRST transformation, we make following ansatz for $S_1$ as
\begin{eqnarray} 
S_1 &=&i\int d^4 x \left[\xi_1(\kappa)\ B (\dot\lambda -\varphi -\vartheta)  +\xi_2(\kappa)\ \dot{\bar c}\dot c
\right.\nonumber\\
&+&\left.\xi_3 (\kappa)\ \bar c c\right],\label{sn1}
\end{eqnarray}
where $\xi_i (i=1, 2, 3)$ are arbitrary $\kappa$-dependent constants and satisfy the
initial conditions $\xi_i (\kappa=0)=0$.

The essential condition in Eq. (\ref{mcond})  satisfies with Eqs. (\ref{jac1}) and (\ref{sn1}) iff
\begin{eqnarray} 
 &&\int d^4x 
\left[- B (\dot\lambda -\varphi -\vartheta)(\xi_1' +\gamma) - \dot{\bar c}\dot c(\xi_2' +\gamma) \right.\nonumber\\
&-&\left.\bar c c(2\gamma -\xi_3') 
+ B\ddot c \Theta_1' (\xi_1 -\xi_2)
 + Bc\Theta_1' (2\xi_1 +\xi_3) \right]\nonumber\\
& =&0,
\end{eqnarray}
where prime denotes the derivative with respect to $\kappa$.
Equating the both sides of the above equation, we get the following  equations 
\begin{eqnarray} 
\xi_1' +\gamma &=&0, \ \ \xi_2' +\gamma =0,\ \ \xi_3' -2\gamma =0,\nonumber\\
 \xi_1 -\xi_2 &=& 0=2\xi_1 +\xi_3. 
\end{eqnarray}
The solution of above equations satisfying the initial conditions is
\begin{eqnarray}
\xi_1 =-\gamma\kappa, \ \ \xi_2 =-\gamma\kappa, \ \ \xi_3 =2\gamma\kappa.
\end{eqnarray}
Then the expression for $S_1$ in terms of $\kappa$ becomes 
\begin{eqnarray}
S_1 =i\int d^4 x \left[-\gamma\kappa B (\dot\lambda -\varphi -\vartheta)  -\gamma\kappa \dot{\bar c}\dot c
+2\gamma\kappa \bar c c\right]. 
\end{eqnarray}
On the other hand the infinitesimal change in Jacobian $J_2$ for the FF-anti-BRST parameter $\Theta_2$  
is calculated as
\begin{equation}
\frac{1}{J_2}\frac{dJ_2}{d\kappa} = - \gamma\int d^4x \left[B (\dot\lambda -\varphi -\vartheta) +\dot{\bar c}\dot c -
2\bar c c\right].
\end{equation}
Similarly, to write the Jacobian $J_2$ as $e^{iS_2}$ in the anti-BRST case, we make ansatz for $S_2$ as
\begin{eqnarray} 
S_2 &=&i\int d^4 x \left[\xi_4(\kappa)\ B (\dot\lambda -\varphi -\vartheta)  +\xi_5(\kappa)\ \dot{\bar c}\dot c
\right.\nonumber\\
&+&\left.\xi_6(\kappa)\ \bar c c\right], 
\end{eqnarray}
where arbitrary $\kappa$-dependent constants $\xi_i  (i=4, 5, 6)$ have to be calculated.

The essential condition in Eq. (\ref{mcond}) for the above Jacobian $J_2$ and functional $S_2$  provides 
\begin{eqnarray}
  \int & d^4x&  
\left[  B (\dot\lambda -\varphi -\vartheta)(\xi_4' -\gamma)+ \dot{\bar c}\dot c(\xi_5' -\gamma) 
-\bar c c(2\gamma +\xi_6') \right.\nonumber\\
&+&\left.B\ddot {\bar c} \Theta_2' (\xi_4 -\xi_5)
 + B\bar c\Theta_2' (2\xi_4 +\xi_6) \right] =0. 
\end{eqnarray}
Comparing the L.H.S. and R.H.S. of the above equation, we get following equations 
\begin{eqnarray}
\xi_4' -\gamma &=&0,\ \xi_5' -\gamma=0,\ \xi_6'+ 2\gamma =0,\nonumber\\
 \xi_4 -\xi_5 &=&0 =2\xi_4 +\xi_6.
\end{eqnarray}
Solving the above equations, we get the following values for $\xi_i$'s 
\begin{eqnarray}
\xi_4 = \gamma\kappa, \ \ \xi_5 = \gamma\kappa, \ \ \xi_6 = -2\gamma\kappa.
\end{eqnarray}
Putting these values in expression of $S_2$, we get
\begin{eqnarray}
S_2 =i\int d^4 x \left[ \gamma\kappa B (\dot\lambda -\varphi -\vartheta)  +\gamma\kappa \dot{\bar c}\dot c
-2\gamma\kappa \bar c c\right]. 
\end{eqnarray}  
Thus under successive FFBRST and FF-anti-BRST transformations the generating functional transformed
as 
\begin{eqnarray}
Z_{CB} \left(  \int D\phi\ e^{iS_{CB}} \right) &&\stackrel{(FFBRST)(FF-anti-BRST)}{--------\longrightarrow } 
Z_{CB}\nonumber\\
&&\left(  \int D\phi\ e^{iS_{CB}+S_1+S_2} \right),
\end{eqnarray}
Note for the particular choices of $\Theta_1$ and $\Theta_2$, the $S_1$ and $S_2$ cancel each other.
Hence $Z_{CB}$ remains invariant under successive FFBRST and FF-anti-BRST transformations.
It is interesting to note that the effect of FMBRST transformation is equivalent to successive operation of
FFBRST and FF-anti-BRST transformations.
 \subsection{Maxwell's theory} 
 The generating functional for Maxwell theory,
using Nakanishi Lautrup type auxiliary field ($B$),
can be given as
\begin{equation}
Z_{M}=\int D\phi\ e^{iS_{eff}^M},\label{zy}
\end{equation}
where the effective action in covariant (Lorentz) gauge with the ghost term is
\begin{equation}
S_{eff}^M=\int d^4 x\left[-\frac{1}{4}F_{\mu\nu}F^{\mu\nu}+\lambda B^2-B\partial_\mu A^\mu
-\bar c\partial_\mu\partial^\mu c\right].
\end{equation}
The infinitesimal off-shell nilpotent BRST and anti-BRST transformations under which the  effective action
$S_{eff}^M$ as well as generating functional $Z_{M}$ remain invariant, are given as

BRST:
\begin{eqnarray}
\delta_b A_\mu &=& \partial_\mu c\ \delta \Lambda_1,\ \ 
\delta_b c =  0\nonumber\\ 
\delta_b \bar c &=& B\ \delta \Lambda_1,\ \ 
\delta_b B = 0.
\end{eqnarray}
Anti-BRST:
\begin{eqnarray}
\delta_{ab} A_\mu &=& \partial_\mu \bar c\ \delta \Lambda_2,\ \ 
\delta_{ab} \bar c = 0,\nonumber\\ 
\delta_{ab} c& = & - B\ \delta \Lambda_2,\ \ 
\delta_{ab} B =  0.
\end{eqnarray}
The nilpotent BRST transformation ($s_b$) and anti-BRST transformation 
($s_{ab}$) mentioned above are absolutely anticommuting in nature i.e.
$\{s_b, s_{ab}\}\equiv s_bs_{ab}+s_{ab}s_b=0$. Therefore
the sum of these two transformations
($s_b$ and $s_{ab}$) is also a nilpotent symmetry transformation. Let us define 
MBRST transformation ($\delta_m \equiv \delta_b+\delta_{ab} $) in this case, which is characterized by 
two infinitesimal parameters $\delta \Lambda_1$ and $\delta \Lambda_2$, as
\begin{eqnarray}
\delta_m A_\mu &=& \partial_\mu c\ \delta \Lambda_1 +\partial_\mu \bar c\ \delta 
\Lambda_2,\nonumber\\
\delta_m c&=& - B\ \delta \Lambda_2,\nonumber\\
\delta_m \bar c &=& B\ \delta \Lambda_1,\nonumber\\
\delta_m B&=& 0.
\end{eqnarray}
The nilpotent FMBRST symmetry transformation for this theory is then
constructed as
\begin{eqnarray}
\delta_m A_\mu &=& \partial_\mu c\ \Theta_1 +\partial_\mu \bar c\ \Theta_2,\nonumber\\
\delta_m c&=& - B\ \Theta_2,\nonumber\\
\delta_m \bar c &=& B\ \Theta_1,\nonumber\\
\delta_m B&=& 0,\label{fin}
\end{eqnarray}
where $\Theta_1$ and $\Theta_2$ are finite, field dependent  and anticommuting parameters.
We choose particular    $\Theta_1$ and $\Theta_2$ in this case as
\begin{equation}
\Theta_1=\int\Theta_1'd\kappa =\gamma\int d\kappa\int d^4x [ \bar c \partial_\mu A^\mu], \label{fi1}
\end{equation}
\begin{equation}
\Theta_2=\int\Theta_2'd\kappa =\gamma\int d\kappa\int d^4x[  c\partial_\mu A^\mu ],\label{fib}
\end{equation} 
where $\gamma$ is an arbitrary parameter. The infinitesimal change in Jacobian 
using Eq. (\ref{jac}) for the FMBRST transformation with the above finite parameters vanishes.   
It means that the path integral measure and hence the generating functional is invariant under
FMBRST transformation.

Now, the infinitesimal change in Jacobian  for the FFBRST transformation with 
the  parameter $\Theta_1$  is calculated as 
\begin{equation}
\frac{1}{J_1}\frac{dJ_1}{d\kappa} =  \gamma\int d^4x \left[B\partial_\mu A^\mu +\bar c \partial_\mu\partial^\mu c
\right].
\end{equation}
To write the Jacobian $J_1$ as $e^{iS_1}$ in the BRST case, we make following ansatz for $S_1$ as
\begin{eqnarray}
S_1 =i\int d^4 x \left[\xi_1 B\partial_\mu A^\mu +\xi_2 \bar c \partial_\mu\partial^\mu c\right].\label{sn}
\end{eqnarray}
The essential condition in Eq. (\ref{mcond}) is satisfied subjected to
\begin{eqnarray} 
 \int &d^4x&
\left[ B\partial_\mu A^\mu (\xi_1' +\gamma) + \bar c \partial_\mu\partial^\mu c (\xi_2' +\gamma) \right.\nonumber\\
&-&\left. B\partial_\mu\partial^\mu c \Theta_1' (\xi_1 -\xi_2) \right] =0,
\end{eqnarray}
where prime denotes the derivative with respect to $\kappa$.
Equating the both sides of the above equation, we get the following   Eqs.  
\begin{eqnarray}
\xi_1' +\gamma =0, \ \ \xi_2' +\gamma =0,\ \ \xi_1 -\xi_2 =0.
\end{eqnarray}
The solution of above equations satisfying the initial conditions $\xi_i =0, (i=1,2)$ is
\begin{eqnarray}
\xi_1 =-\gamma\kappa, \ \ \xi_2 =-\gamma\kappa.
\end{eqnarray}
Putting these value  in the Eq. (\ref{sn}), the expression of $S_1$ becomes 
\begin{eqnarray}
S_1 =-i\gamma\kappa\int d^4 x \left[  B\partial_\mu A^\mu + \bar c \partial_\mu\partial^\mu c\right].\label{s1}
\end{eqnarray}
However, the infinitesimal change in Jacobian $J_2$ for the FF-anti-BRST transformation with 
the parameter $\Theta_2$  
is calculated as
\begin{equation}
\frac{1}{J_2}\frac{dJ_2}{d\kappa} = -\gamma\int d^4x \left[B\partial_\mu A^\mu +\bar c \partial_\mu\partial^\mu c
\right].\label{j22}
\end{equation}
Similarly, to write the Jacobian $J_2$ as $e^{iS_2}$ in the anti-BRST case, we make the following ansatz for $S_2$ as
\begin{eqnarray}
S_2 =i\int d^4 x \left[\xi_3 B\partial_\mu A^\mu +\xi_4 \bar c \partial_\mu\partial^\mu c\right].\label{s22}  
\end{eqnarray}
The essential condition in Eq. (\ref{mcond}) for Eqs. (\ref{j22}) and (\ref{s22})  provides
\begin{eqnarray} 
 \int &d^4x&
\left[ B\partial_\mu A^\mu (\xi_3' -\gamma) + \bar c \partial_\mu\partial^\mu c (\xi_4' -\gamma) \right.\nonumber\\
&-&\left. B\partial_\mu\partial^\mu \bar c \Theta_2' (\xi_3 -\xi_4) \right] =0.
\end{eqnarray}
Comparing the L.H.S. and R.H.S. of the above equation we get following equations
\begin{equation}
\xi_3' -\gamma =0,\ \xi_4' -\gamma =0,\ \xi_3 -\xi_4 =0.
\end{equation}
Solving the above equations, we get the following values for $\xi$'s 
\begin{eqnarray}
\xi_3 = \gamma\kappa, \ \ \xi_4 = \gamma\kappa.
\end{eqnarray}
Plugging back these value of $\xi_i (i=3, 4)$ in Eq. (\ref{s22}), we obtain   
\begin{eqnarray}
S_2 = i\gamma\kappa\int d^4 x \left[  B\partial_\mu A^\mu + \bar c \partial_\mu\partial^\mu c\right].\label{s2}
\end{eqnarray}
From Eqs. (\ref{s1}) and (\ref{s2}), one can easily see that 
$S_1+S_2=0$.
Therefore, under successive FFBRST and FF-anti-BRST transformations with these particular finite parameters 
$\Theta_1$and  $\Theta_2$ respectively, the generating functional 
transformed
as  
\begin{eqnarray}
Z_M \left(  \int D\phi\ e^{iS_{eff}^M} \right)&&\stackrel{(FFBRST)(FF-anti-BRST)}{--------\longrightarrow }
Z_M\nonumber\\
&&\left(   \int D\phi\ e^{iS_{eff}^M+S_1+S_2} \right).
\end{eqnarray}
Hence the successive operation of FFBRST and FF-anti-BRST transformation also  leaves the generating functional 
$Z_M$.  This reconfirms that FMBRST transformation has same effect on both  the effective action and  the generating 
functional as the successive FFBRST and FF-anti-BRST transformations.  
\subsection{Non-Abelian YM theory in CFDJ gauge}
The generating functional for non-Abelian YM theory in CFDJ gauge can be written as
\begin{equation}
Z^{CF}_{YM } =\int D\phi\ e^{iS^{CF}_{YM }[\phi] },\label{zy1}
\end{equation}
where $\phi$ is generic notation for all the fields in the effective action 
$S^{CF}_{YM }$  
\begin{eqnarray}  
&&S^{CF}_{YM } = \int d^4x \left[ -\frac{1}{4}F_{\mu\nu}^aF^{\mu\nu a}+\frac{\xi}{2}(h^a)^2 
+ih^a
\partial_\mu A^{\mu a}\right.\nonumber\\
 &+&\left. \frac{1}{2}\partial_\mu{\bar c}^a(D^\mu c)^a +\frac{1}{2}(D_\mu\bar 
c)^a\partial^\mu c^a
-\xi\frac{g^2}{8}
 (f^{abc}{\bar c}^b c^c)^2\right] 
\end{eqnarray}
with the  field strength tensor $F_{\mu\nu}^a=\partial_\mu A_\nu^a -\partial_\nu A_\mu^a +gf^{abc}A_\mu^b A_\nu^c$ 
and 
$h^a$ is the Nakanishi Lautrup type auxiliary field. 
 The effective action as well as the generating functional are invariant under following infinitesimal BRST 
and anti-BRST transformations 

BRST: \begin{eqnarray}
\delta_b A_\mu^a&=& -(D_\mu c)^a \ \delta\Lambda_1, \ \ \ \delta_b c^a=-\frac{g}{2}f^{abc}{ c}^b c^c \ 
\delta\Lambda_1
 \nonumber\\
\delta_b {\bar c}^a &=& \left(ih^a -\frac{g}{2}f^{abc}{\bar c}^b c^c\right)\ \delta
\Lambda_1\nonumber\\
\delta_b (ih^a)&=& -\frac{g}{2}f^{abc}\left( ih^b c^c  +\frac{g}{4} f^{cde}{\bar c}^b c^d c^e\right)
\ \delta\Lambda_1,
\end{eqnarray}
Anti-BRST:
\begin{eqnarray}
\delta_{ab} A_\mu^a&=& -(D_\mu \bar c)^a \ \delta\Lambda_2, \ \ \ \delta_{ab} {\bar c}^a=
-\frac{g}{2}f^{abc}{\bar c}^b {\bar c}^c \ \delta\Lambda_2
 \nonumber\\
\delta_{ab} { c}^a &=& \left(-ih^a -\frac{g}{2}f^{abc}{\bar c}^b c^c\right)\ 
\delta\Lambda_2\nonumber\\
\delta_{ab} (ih^a)&=& -\frac{g}{2}f^{abc}\left( ih^b {\bar c}^c  +\frac{g}{4} f^{cde}{ c}^b {\bar 
c}^d {\bar c}^e\right)
\ \delta\Lambda_2,
\end{eqnarray}
where $\delta\Lambda_1$ and $\delta\Lambda_2$ are infinitesimal, anticommuting and global parameters.
The infinitesimal MBRST symmetry transformation ($\delta_m= \delta_b +\delta_{ab}$) in this case is written as:
\begin{eqnarray} 
\delta_m A_\mu^a&=& -D_\mu c^a\ \delta\Lambda_1 - D_\mu \bar c^a \ \delta\Lambda_2,\nonumber\\ 
\delta_m c^a&=&  -\frac{g}{2}f^{abc}{ c}^b c^c \ 
\delta\Lambda_1 -  \left(ih^a +\frac{g}{2}f^{abc}{\bar c}^b c^c\right)\ 
\delta\Lambda_2,
 \nonumber\\
\delta_m {\bar c}^a &=&  \left(ih^a -\frac{g}{2}f^{abc}{\bar c}^b c^c\right)\ \delta
\Lambda_1  
-\frac{g}{2}f^{abc}{\bar c}^b {\bar c}^c \ \delta\Lambda_2,\nonumber\\
\delta_m (ih^a)&=& -\frac{g}{2}f^{abc}\left( ih^b c^c  +\frac{g}{4} f^{cde}{\bar c}^b c^d c^e
\right)
\ \delta\Lambda_1 \nonumber\\
&-&\frac{g}{2}f^{abc}\left( ih^b {\bar c}^c  +\frac{g}{4} f^{cde}{ c}^b {
\bar 
c}^d {\bar c}^e\right)
\delta\Lambda_2. 
\end{eqnarray} 
Corresponding FMBRST symmetry transformation 
is constructed as:
\begin{eqnarray}
\delta_m A_\mu^a&=& -D_\mu c^a\ \Theta_1 - D_\mu \bar c^a \ \Theta_2,\nonumber\\
 \delta_m c^a &=&  -\frac
{g}{2}f^{abc}{ c}^b c^c \ 
\Theta_1 -  \left(ih^a +\frac{g}{2}f^{abc}{\bar c}^b c^c\right)\ 
\Theta_2,
 \nonumber\\
\delta_m {\bar c}^a &=&  \left(ih^a -\frac{g}{2}f^{abc}{\bar c}^b c^c\right)\
\Theta_1  
-\frac{g}{2}f^{abc}{\bar c}^b {\bar c}^c \ \Theta_2,\nonumber\\
\delta_m (ih^a)&=& -\frac{g}{2}f^{abc}\left( ih^b c^c  +\frac{g}{4} f^{cde}{\bar c}^b c^d c^e
\right)
\ \Theta_1\nonumber\\ 
&-&\frac{g}{2}f^{abc}\left( ih^b {\bar c}^c  +\frac{g}{4} f^{cde}{ c}^b {\bar 
c}^d {\bar c}^e\right)
\ \Theta_2,\label{fini}
\end{eqnarray}
with two arbitrary finite field dependent parameters  $\Theta_1$ and $\Theta_2$.
The generating functional $Z^{CF}_{YM }$ is made invariant under the above FMBRST
transformation  by constructing appropriate finite parameters  $\Theta_1$ and $\Theta_2$.
We construct  the  finite nilpotent  parameters $\Theta_1$ and $\Theta_2$ 
 as
\begin{equation}
\Theta_1 =\int \Theta_1' d\kappa =\gamma \int d\kappa \int d^4x[\bar c^a\partial_\mu A^{\mu a}], \label{fi2}
\end{equation}
 \begin{equation}
\Theta_2 =\int \Theta_2' d\kappa =\gamma \int d\kappa \int d^4 x[  c^a\partial_\mu A^{\mu a}], \label{fic}
\end{equation} 
where $\gamma$ is an arbitrary parameter.
Following the same method elaborated in previous two examples, we show that the Jacobian for path  integral 
measure due to FMBRST transformation given in Eq. (\ref{fini}) 
with finite parameters  $\Theta_1$ and $\Theta_2$ becomes unit.
It means that under such FMBRST transformation the generating functional as well as effective action remain
invariant. The Jacobian contribution for path integral measure due to FFBRST 
transformation with parameter $\Theta_1$ compensates 
the same due to FF-anti-BRST transformation with parameter $\Theta_2$.
Therefore, under the successive FFBRST and FF-anti-BRST transformations the 
generating functional remains invariant as
\begin{eqnarray} 
Z_{YM}^{CF}  \stackrel{(FFBRST)(FF-anti-BRST)}{---------\longrightarrow }
Z_{YM}^{CF}.
\end{eqnarray}
Again we see the equivalence between FMBRST and successive operation of FFBRST and FF-anti-BRST transformations.
  
We end up the section with conclusion that in all the three cases the FMBRST transformation with appropriate
finite parameters is the finite nilpotent symmetry of the effective action as well as the 
generating functional of the effective theories. Here we also note that the successive operations of FFBRST and
FF-anti-BRST  also leave the generating functional  as well as effective action invariant and hence equivalent 
to FMBRST transformation. 
\section{FMBRST symmetry in field/antifield formulation} 
 In this section we consider the field/antifield formulation using MBRST transformation.
 Unlike BV formulation using either BRST or anti-BRST transformations,
 we need two sets of antifields in BV formulation for MBRST transformation. We 
construct FMBRST transformation in this context.
 The change in Jacobian under FFBRST transformation in the path integral measure in the definition of  
generating functional is used to adjust with the change in the gauge-fixing fermion $\Psi_1$ \cite{sudbpm}.
Hence the FFBRST  transformation is used to connect the generating functionals of different solutions 
of quantum master equation \cite{srbpm,subp}. However in case of  BV formulation for FMBRST transformation
 we need to introduce two gauge-fixing fer-\\
mions  $\Psi_1$ and $\Psi_2$.
 We construct the finite parameters in FMBRST transformation in such a way that contributions from
 $\Psi_1$ and $\Psi_2$ adjust each other  to leave the extended action invariant. This implies that we can construct 
appropriate parameters in FMBRST transformation such that generating functionals corresponding to different 
solutions of quantum master equations remain invariant under such transformation. 
These results can be demonstrated with the help of 
explicit examples. We would like to consider the same examples of previous section for this purpose. 
\subsection{Bosonized chiral model in BV formulation}
We recast the generating functional in Eq. (\ref{z}) for (1+1) dimensional bosonized chiral model 
using both BRST and anti-BRST exact terms as
\begin{eqnarray}  
Z_{CB}&=&\int D\phi\ e^{iS_{CB}},\nonumber\\
& =& \int D\phi\ \exp \left[i\int d^2 x\left\{\pi_\varphi\dot\varphi +
\pi_\vartheta\dot\vartheta
+p_u\dot u -\frac{1}{2}\pi_\varphi^2 \right.\right.\nonumber\\
&+& \left.\left.\frac{1}{2}\pi_\vartheta^2 
+\pi_\vartheta (\varphi'-\vartheta' +\lambda ) +\frac{1}{2}\pi_\varphi \lambda  +
\frac{1}{2}s_b \Psi_1\right.\right.\nonumber\\
& +&\left.\left.\frac{1}{2}s_{ab}\Psi_2\right\}\right],\label{ac} 
\end{eqnarray}
here Lagrange multiplier field $u$ is considered as dynamical variable and 
expression for gauge-fixing fermions for BRST symmetry ($\Psi_1 $) and   anti-BRST symmetry ($\Psi_2$) 
respectively are
\begin{eqnarray}
\Psi_1 &=&\int d^2x\ \bar c(\dot\lambda-\varphi-\vartheta +\frac{1}{2} B).\\
\Psi_2 &=&\int d^2x\  c (\dot\lambda-\varphi-\vartheta +\frac{1}{2} B).
\end{eqnarray}
The effective  action $S_{CB}$ is invariant under combined BRST and anti-BRST 
transformations given in Eq. (\ref{b}).  
The generating functional $Z_{CB}$ can be written in terms of antifields  $\phi_1^\star$ and $\phi_2^\star$ 
corresponding to
all fields $\phi$ as
\begin{eqnarray} 
Z_{CB}&=& \int D\phi\ \exp \left[i\int d^2 x\left\{\pi_\varphi\dot\varphi +\pi_\vartheta\dot\vartheta
+p_u\dot u -\frac{1}{2}\pi_\varphi^2 \right.\right.\nonumber\\
&+& \left.\left.\frac{1}{2}\pi_\vartheta^2 +\pi_\vartheta (\varphi'-\vartheta' +\lambda ) 
+\pi_\varphi \lambda +\frac{1}{2}
\varphi_1^\star
c  -\frac{1}{2}
\varphi_2^\star
 \bar c  \right.\right.\nonumber\\
&+&\left.\left. \frac{1}{2}
\vartheta_1^\star   c  -\frac{1}{2} \vartheta_2^\star
\bar c  +\frac{1}{2}\bar c_1^\star  B +\frac{1}{2}c_2^\star B 
- \frac{1}{2}\lambda_1^\star 
\dot c\right.\right.\nonumber\\
&+&\left.\left.\frac{1}{2} \lambda_2^\star 
\dot {\bar c}\right\}\right],
\end{eqnarray}
where $\phi_i^\star (i=1,2)$ is a generic notation for antifields arising from gauge-fixing fermions 
$\Psi_i$.
The above relation can further be written in compact form as 
\begin{equation}
Z_{CB} = \int D\phi\  e^{ iW_{\Psi_1 +\Psi_2}[\phi,\phi_i^\star ] },\label{cq}
\end{equation}
where $W_{\Psi_1 +\Psi_2} [\phi,\phi_i^\star ]$ is an extended action for the theory of self-dual chiral 
boson corresponding the gauge-fixing fermions $\Psi_1$ and $\Psi_2$. 

This extended quantum action, $W_{\Psi_1 +\Psi_2} [\phi,\phi_i^\star ]$ satisfies certain rich 
mathematical
 relations commonly known as quantum master equation \cite{wei}, given by
\begin{equation}
\Delta e^{iW_{\Psi_1 +\Psi_2}[\phi,\phi_i^\star ]} =0  \ \mbox{ with }\ 
 \Delta\equiv \frac{\partial_r}{
\partial\phi}\frac{\partial_r}{\partial\phi_i^\star } (-1)^{\epsilon
+1}. 
\label{mq}
\end{equation}
The generating functional does not depend on the choice of gauge-fixing fermions \cite{ht} and therefore  
extended quantum action $W_{\Psi_i}$  with all possible $\Psi_i$  are the different 
solutions of quantum master equation.
The antifields $\phi^\star_1$ corresponding to each field $\phi$ for this particular theory 
can 
be obtained from the gauge-fixed fermion $\Psi_1$ as
\begin{eqnarray}
\varphi_1^\star &=& \frac{\delta \Psi_1}{\delta \varphi}= - \bar c,
\ \ \vartheta_1^\star = \frac{\delta \Psi_1}{\delta \vartheta}= - \bar c,
{ c}_1^\star =\frac{\delta \Psi_1}{\delta c}=0,\nonumber\\
 { \bar c}_1^\star &=&  \frac{\delta 
\Psi_1}{\delta  \bar c}=-\frac{1}{2}(
\dot\lambda -\varphi- \vartheta +\frac{1}{2}B),\nonumber\\
B_1^\star &=& \frac{\delta \Psi_1}{\delta B}= \frac{1}{2}\bar c,\ \ \lambda_1^\star =-\frac{
\delta \Psi_1}{\delta \lambda}= - \dot{\bar c}.
\end{eqnarray}
Similarly, the  antifields $\phi^\star_2$ can be
  calculated from the gauge-fixing fermion $\Psi_2$ as
\begin{eqnarray}
\varphi_2^\star &=& \frac{\delta \Psi_2}{\delta \varphi}= -  c,
\ \ \vartheta_2^\star = \frac{\delta \Psi_2}{\delta \vartheta}= -  c,\nonumber\\
{ c}_2^\star &=& \frac{\delta \Psi_2}{\delta  c}= (
\dot\lambda -\varphi- \vartheta +\frac{1}{2}B),\ \ {\bar c}_2^\star = \frac{\delta \Psi_2}{
\delta \bar c}=0,\nonumber\\
B_2^\star &=& \frac{\delta \Psi_2}{\delta B}=\frac{1}{2}   c,\ \ \lambda_2^\star =-\frac{
\delta \Psi_2}{\delta \lambda}= -  \dot {  c}.
\end{eqnarray}
Now we apply the FMBRST transformation given in Eq. (\ref{c}) with the finite parameters written in Eqs. (\ref{fi}) 
and (\ref{fia}) to this generating functional.
We see that the path integral measure in Eq. (\ref{cq}) remains invariant under this FMBRST transformation
as the Jacobian for path integral measure is $1$.
Therefore,
\begin{eqnarray}
Z_{CB}\left( \int D\phi\ e^{i W_{\Psi_1 +\Psi_2}}\right)  &&\stackrel{(FFBRST)(FF-anti-BRST)}{ -------
\longrightarrow}\nonumber\\  
 &&Z_{CB}. 
    \end{eqnarray}
Thus the solutions of quantum master equation in this model remain invariant under FMBRST
 transformation  as well as under consecutive operations of FFBRST and FF-anti-BRST transformations. 
However  the FFBRST (FF-anti-BRST) transformation connects the generating functionals corresponding to
the different solutions of the quantum master equation \cite{srbpm,ssb}.
\subsection{Maxwell's theory in BV formulation}
The generating functional for Maxwell's theory given in Eq. (\ref{zy}) can be  recast 
using BRST and anti-BRST exact terms as  
\begin{eqnarray}
Z_{M}= \int D\phi\  e^{ i\int d^4 x\left[ -\frac{1}{4}F_{\mu\nu}F^{\mu\nu}+ \frac{1}{2}s_b \Psi_1 +
\frac{1}{2}s_{ab} \Psi_2 \right]},
\end{eqnarray}
where the  expressions for gauge-fixing fermions $\Psi_1 $ and $\Psi_2 $ are
\begin{equation}
\Psi_1 = \int d^4x\ \bar c(\lambda B-\partial\cdot A), 
\end{equation}
\begin{equation} 
\Psi_2 =-\int d^4x\ c (\lambda B-\partial\cdot A).
\end{equation}

The generating functional for such theory can  further be expressed in fields/antifields formulation as 
\begin{eqnarray}
Z_{M}&=& \int D\phi\  \exp \left[ i\int d^4 x\left( -\frac{1}{4}F_{\mu\nu}F^{\mu\nu}+\frac{1}{2}A_{ \mu 1}^\star 
 \partial^\mu c\right.\right.\nonumber\\
&+&\left.\left.\frac{1}{2} A_{\mu 2}^\star\partial^\mu \bar c +\frac{1}{2}{\bar c}_1^\star B-\frac{1}{2}
{ c}_2^\star B \right)\right].
\end{eqnarray} 
In the compact form above generating functional is written as
\begin{equation}
 Z_{M} = \int D\phi e^{iW_{\Psi_1+\Psi_2} [\phi,\phi_i^\star ]},
\end{equation} 
 where $W_{\Psi_1 +\Psi_2}[\phi,\phi_i^\star ]$ is an extended action for the Maxwell's theory corresponding to
 the gauge-fixing fermions $\Psi_1 $ and $\Psi_2$.

The antifields for gauge-fixed fermion $\Psi_1$ are calculated as
\begin{eqnarray}
A_{ \mu 1}^{\star} &=&\frac{\delta \Psi_1}{\delta A^\mu}= 
\partial_\mu {\bar c}, \ \ \bar c_1^{\star} =\frac{\delta \Psi_1}{\delta \bar c}=
(\lambda B-\partial\cdot A),\nonumber\\
{ c}_1^{\star} &=&\frac{\delta \Psi_1}{\delta {c}}=0,
\ \ B_1^\star =\frac{\delta \Psi_1}{\delta B}=\lambda  \bar c.
\end{eqnarray}
The antifields $\phi_2^\star$ can 
be calculated from the gauge-fixed fermion $\Psi_2$ as 
\begin{eqnarray}
A_{ \mu 2}^{\star} &=&\frac{\delta \Psi_2}{\delta A^\mu}= - 
\partial_\mu { c}, \ \ \bar c_2^{\star} =\frac{\delta \Psi_2}{\delta \bar c}=0,\nonumber\\
{ c}_2^{\star} &=&\frac{\delta \Psi_2}{\delta {c}}=-(\lambda B-\partial\cdot A),
\  B_2^\star =\frac{\delta \Psi_2}{\delta B}= -\lambda c.
\end{eqnarray}
 Now implementing the FMBRST transformation mentioned in Eq. (\ref{fin}) with parameters given in Eqs. (\ref{fi1})
and (\ref{fib}) to this generating functional  we see that the Jacobian  for path integral
measure for such transformation becomes unit. Hence, the FMBRST transformation given in Eq. (\ref{fin}) is a finite 
symmetry of the solutions of quantum master equation for Maxwell's theory.

Now, we focus on the contributions arising from the Jacobian due to independent applications of FFBRST and 
FF-anti-BRST transformations. We construct the finite parameters of FFBRST and FF-anti-BRST transformations
in such a way that Jacobian remains invariant.
Therefore, the generating functional remains invariant under consecutive operation of FFBRST
and FF-anti-BRST transformations with appropriate parameters as 
\begin{eqnarray}
Z_{M}\stackrel{(FFBRST)(FF-anti-BRST)}{---------\longrightarrow} 
 Z_{M}.
    \end{eqnarray}
It also implies that the effect of consecutive FFBRST and FF-anti-BRST transformations is same as the effect 
of FMBRST transformation on $Z_M$. 
\subsection{Non-Abelian YM theory in BV formulation}
  The generating functional for this theory can be written 
in both BRST and anti-BRST exact terms as  
\begin{eqnarray}
Z^{CF}_{YM } =  \int D\phi\ e^{i\int d^4 x\left[ -\frac{1}{4}F_{\mu\nu}^a
F^{a\mu\nu}+ \frac{1}{2}s_b \Psi_1 +\frac{1}{2}s_{ab} \Psi_2 \right ]},
\end{eqnarray}
with the expressions of gauge-fixing fermions  $\Psi_1$ and $\Psi_2$ as
\begin{equation}
\Psi_1 =-\int d^4x\ \bar c^a(i\frac{\xi}{2}h^a-\partial\cdot A^a),
\end{equation} 
\begin{equation}
\Psi_2 =\int d^4x\ c^a (i\frac{\xi}{2}h^a-\partial\cdot A^a).
\end{equation}
We re-write the generating functional given in Eq. (\ref{zy1}) using field/antifield formulation   as,
\begin{eqnarray}   
 Z^{CF}_{YM } &=&  \int D\phi\ \exp \left[i\int d^4 x\left\{- \frac{1}{4}F_{\mu\nu}^a
F^{a\mu\nu}-\frac{1}{2}A_{ \mu 2}^{a\star} 
D^\mu c^a\right.\right.\nonumber\\
&-&\left.\left.\frac{1}{2} A_{ \mu 2}^{a\star}D^\mu \bar c^a +\frac{1}{2}{\bar c}_1^{a\star} \left(
ih^a -\frac{g}{2}f^{abc}\bar c^b c^c\right)\right.\right.\nonumber\\
&-&\left.\left. \frac{1}{2}{ c}_2^{a\star} \left(
ih^a +\frac{g}{2}f^{abc}\bar c^b c^c\right)
-\frac{1}{2} 
h_1^{a\star}\left( \frac{g}{2}f^{abc}h^b c^c \right.\right.\right.\nonumber\\
&- &\left.\left.\left. i \frac{g^2}{8}f^{abc}f^{cde}\bar c^b{  c}^d{c}^e \right)
-
\frac{1}{2} 
h_2^{a\star}\left( \frac{g}{2}f^{abc}h^b \bar c^c \right.\right.\right.\nonumber\\
&- &\left.\left.\left. i \frac{g^2}{8}f^{abc}f^{cde} c^b{\bar c}^d{\bar c}^e 
\right)\right\}\right].
\end{eqnarray}
This generating functional $Z^{CF}_{YM }$ can be written compactly as
\begin{equation}
Z^{CF}_{YM } = \int D\phi\ e^{  iW_{\Psi_1+\Psi_2}[\phi,\phi_i^\star ]},
\end{equation}
where $W_{\Psi_1+\Psi_2}[\phi,\phi_i^\star  ]$ is an extended quantum action for the non-Abelian YM theory 
in CFDJ gauge.

The antifields   are calculated with the help of gauge-fixed fermion $\Psi_1$ as
\begin{eqnarray} 
A_{ \mu 1}^{a\star} &=&\frac{\delta \Psi_1}{\delta A^{a\mu}}=-\partial_\mu {\bar c}^a,
\ \ {  c}_1^{a\star}  = \frac{\delta \Psi_1}{\delta {  c^a}}=0,\nonumber\\
 \bar c_1^{a\star} &=&\frac{\delta \Psi_1}{\delta \bar c^a}=- 
(i\frac{\xi}{2}h^a-\partial\cdot A^a),\nonumber\\
  h_1^{a\star} &=&\frac{\delta \Psi_1}{\delta h^a}=
-\frac{i}{2}\xi \bar c^a.
\end{eqnarray}
The explicit value of antifields  can be calculated with   $\Psi_2$ as  
\begin{eqnarray}
A_{ \mu 2}^{a\star} &=&\frac{\delta \Psi_2}{\delta A^{a\mu}} =\partial_\mu { 
c^a}, \ \ \bar c_2^{a\star} =\frac{\delta \Psi_2}{\delta \bar c^a}=0,\nonumber\\
{  c}_2^{a\star} &=&\frac{\delta \Psi_2}{\delta { c^a}}= 
(i\frac{\xi}{2}h^a-\partial\cdot A^a),\nonumber\\
 h_2^{a\star} &=&\frac{\delta \Psi_2}{\delta h^a}=
\frac{i}{2}\xi {  c}^a.
\end{eqnarray}
We observe here again that the Jacobian for path integral measure in the expression of generating functional 
$Z_{YM}^{CF}$ arising due to FMBRST transformation and 
 due to successive operation of FFBRST and FF-anti-BRST transformations remains unit
for appropriate choice of finite parameters. 
Thus, the consequence of FMBRST transformation given in Eq. (\ref{fini}) with the finite parameters given in Eqs. 
(\ref{fi2}) and (\ref{fic}) is equivalent to the subsequent operations of FFBRST and FF-anti-BRST transformations
with same finite parameters.
\section{Concluding remarks}
FFBRST and FF-anti-BRST transformations are nilpotent symmetries of the effective action. However these 
transformations do not leave the generating functional invariant as the path integral measure 
changes in a nontrivial way under these transformations. 
We have constructed infinitesimal MBRST transformation which is the combination of 
infinitesimal BRST and anti-BRST transformations.
Even though infinitesimal MBRST transformation does not play much significant role, its field dependent version
has very important  consequences. We have shown that it is possible to construct the  field dependent
 MBRST (FMBRST) transformation  which leaves the effective action as well as
the generating functional invariant. The finite parameters in the FMBRST transformation have been chosen in such a 
way that the Jacobian contribution from the FFBRST part compensates the same arising from FF-anti-BRST part.
We have considered several explicit examples with diverse character in both gauge theories
as well as in field/antifield formulation to show these results.
It is interesting to point out that the effect of FMBRST transformation is equivalent to successive operations of
FFBRST and FF-anti-BRST transformations. We have further shown that the generating functionals 
corresponding to different solutions of quantum
master equation remain invariant under such FMBRST transformation whereas the independent FFBRST and FF-anti-BRST
 transformations  connect the generating functionals corresponding to
the different solutions of the quantum master equation. It will be interesting to see whether these 
FMBRST transformations put further restrictions on the relation of different Green's function of the
theory to simplify the renormalization program. In particular, such FMBRST transformations may be helpful 
for the theories where BRST and anti-BRST transformations play independent role.

\section*{Acknowledgments}
 We thankfully acknowledge the financial support from the Department of Science and Technology 
(DST), Government of India, under the SERC project sanction grant No. SR/S2/HEP-29/2007.

\end{document}